\documentclass[aps,preprint,amssymb]{revtex4}
\usepackage{pifont}
\usepackage{stmaryrd}

\usepackage{epsfig}

\begin{document}
\renewcommand{\thefootnote}{\fnsymbol{footnote}}
\renewcommand{\theequation}{\arabic{section}.\arabic{equation}}

\title{Spectroscopic Evidence for the Specific Na$^{+}$ and K$^{+}$ Interactions with the Hydrogen-bonded Water Molecules at the
Electrolyte Aqueous Solution Surfaces}

\author{Ran-ran Feng\footnote[2]{Also Graduate University of the Chinese Academy of
Sciences.}, Hong-tao Bian\footnotemark[2], Yuan Guo and Hong-fei
Wang\footnote[1]{To whom correspondence should be addressed. E-mail:
hongfei@iccas.ac.cn; Tel: 86-10-62555347; Fax:86-10-62563167.}}


\affiliation{Beijing National Laboratory for Molecular Sciences,
State Key Laboratory of Molecular Reaction Dynamics, Institute of
Chemistry, Chinese Academy of Sciences, Beijing, China 100190}
\date{\today}

\begin{abstract}
\noindent Sum frequency generation vibrational spectra of the water
molecules at the NaF and KF aqueous solution surfaces showed
significantly different spectral features and different
concentration dependence. This result is the first direct
observation of the cation effects of the simple alkali cations,
which have been believed to be depleted from the aqueous surface, on
the hydrogen bonding structure of the water molecules at the
electrolyte solution surfaces. These observations may provide
important clue to understand the fundamental phenomenon of ions at
the air/water interface.
\end{abstract}

\maketitle

In this letter we would like to report the significantly different
sum-frequency generation vibrational spectra of the NaF and KF
aqueous solution surfaces. These results suggested that the ion
effects on the interfacial water hydrogen bonding structure are
quite different for the NaF and KF salts. These results may provide
direct counterexamples to the physical picture based on the
currently prevalent molecular dynamics (MD) simulations with the
polarizable force field, which have predicted that the adsorption of
the more polarizable anions, such as the I$^{-}$ and Br$^{-}$
anions, are significantly enhanced at the electrolyte aqueous
solution surfaces, while the less polarizable anions, such as the
F$^{-}$ and Cl$^{-}$ anions, and the non polarizable cations, such
as the Na$^{+}$ and K$^{+}$ cations, are repelled from the
electrolyte aqueous solution
surfaces.\cite{TobiasChemRev,TobiasFeatureArticle6361,Tobias7617,DangChemRev,TobiasJPCB10468}

Recently, there have been a great deal of theoretical and
experimental studies on the ion solvation and ion effects at the
electrolyte aqueous solution
surfaces.\cite{AllenChemRev,GarrettScience,TobiasScience08,TobiasFeatureArticle6361,
TobiasChemRev,JungwirthARPC,JungwirthFaraday2009,SaykallyCPL51,
SaykallyCPL46,Saykally10915,Saykally7976,SaykallyJACS,SaykallyARPS,
AllenSodiumHalide2252,ShultzJPCB585,Motschmann4484,Motschmann2099,
AllenJPCB8861,ShenJACS13033,BianPCCP,Richmond5051,SloutskinJCP054704,
GhosalScience,WinterChemRev2006,HemmingerPCCP4778,PadmanabhanPRL086105}
Ions have long been considered to be depleted from the surface of an
aqueous solution of simple electrolytes, such as the alkali
halides.\cite{LangmuirJACS1848,OnsagerJCP528,AdamsonBook} According
to the Gibbs adsorption equation, the increase of the surface
tension of water with the addition of simple inorganic salts was
explained by the assertion that the simple ions can only have a
¡®¡®negative adsorption¡¯¡¯ at the aqueous
interface.\cite{AdamsonBook} The subsequent electrostatic theory
incorporated with the image charge repulsion effect, i.e. the
Wagner-Onsager-Samaras model, provided a quantitative molecular
theory for the ion (both anion and cation) depletion picture for the
aqueous interface at low electrolyte
concentration.\cite{OnsagerJCP528,MarkinJPCB11810} Such a classical
picture remained unchallenged till the 1990s.

The excitement to go beyond Onsager and Langmuir certainly ignited
many activities on this seemingly long-settled
problem.\cite{SaykallyARPS,TobiasFeatureArticle6361} Recent
experimental observations on the chemical reaction occurring on
aqueous sea-salt particles and ocean surfaces also indicated a
likely incompatibility with the picture of an ion-free aqueous
surface.\cite{KinppingScience301,LaskinScience340,HuJPC8768}
Following up these experimental observations, recent molecular
dynamics (MD) simulations on the aqueous systems, especially the
development of the polarizable force fields of liquid water
molecules and the anions, predicted specific anion effects at the
interface of simple electrolyte aqueous
solutions.\cite{TobiasChemRev,DangChemRev,TobiasFeatureArticle6361}
These studies predicted the enrichment effect of the halide anions
in the aqueous solution surface region, and also pointed out that
the enrichment was associated with the electrostatic polarizability
of the specific anions.\cite{TobiasJPCB10468} In this paradigm, the
F$^{-}$ and Cl$^{-}$ anions are less polarizable, and the Na$^{+}$
and K$^{+}$ cations are considered non-polarizable. All of them are
expected to be repelled from the aqueous solution surface. Thus, no
or little anion and cation effect on the interfacial water hydrogen
bonding structure by either of these four ions was expected.

Sum-frequency generation vibrational spectroscopy (SFG-VS) has been
proven to be the only surface specific spectroscopic method to probe
the vibrational spectra of the molecular species at the liquid
surfaces.\cite{ShenYRJPCB3292} Since it was first applied to probe
the water molecules at the neat air/water interface by Shen and
co-workers in 1993,\cite{ShenYRPRL2313} tremendous advances in the
understanding of the aqueous surfaces have been
achieved.\cite{ShultzIRPC123,RichmondChemRev2002,ShenChemRev2006,GanWeiJCP114705}
In the SFG-VS experiment, usually a tunable infrared (IR) laser beam
and a fixed wavelength visible beam are focused simultaneously onto
the molecular surface, and the resulted SFG signal at the sum of the
IR and visible frequencies is
detected.\cite{shen1989nature,ShenZhuangPRB,WangIRPC} The ssp
polarization combination is one of the two often used polarization
combinations, i.e. ssp and ppp, in the SFG-VS experiment. Here ssp
means that in the SFG-VS experiment, the electric field vectors of
the outgoing sum frequency signal beam, the incident visible beam
and the incident infrared (IR) beam are in the s, s and p
polarizations, respectively. s represents the case when the electric
field vector of the optical beam is perpendicular to the incident
plan formed by the surface normal vector and the incident beam wave
vectors; while p represents the case when the electric field vector
is in the incident plane.\cite{ShenZhuangPRB}

It has been well established that the ssp SFG spectra for the neat
air/water interface has two prominent features, one is the narrow
peak centered at 3700cm$^{-1}$ and the other is the fairly broad
band in the 3000-3600cm$^{-1}$. The former is the signature of the
non hydrogen-bonded O-H group which sticks out of the liquid phase,
i.e. the so called `free O-H'; while the latter is the signature of
the O-H stretching vibrations of the differently hydrogen-bonded
water molecules on the liquid side in the vicinity of the air/water
interface.\cite{ShenYRPRL2313,ShultzIRPC123,RichmondChemRev2002,ShenChemRev2006,GanWeiJCP114705}
These signatures of the interfacial water molecules can be observed
in the SFG-VS because the water molecules at the air/water interface
are orientationally
ordered.\cite{ShenYRPRL2313,ShultzIRPC123,RichmondChemRev2002,ShenChemRev2006,GanWeiJCP114705,GanWeiCJCP20}

It has been known that addition of the electrolytes into the bulk
aqueous phase can introduce changes in the ssp SFG spectra in the
hydrogen-bonded water region, i.e.
3000-3600cm$^{-1}$.\cite{ShultzIRPC123,ShultzJPCBFeature5313,
AllenChemRev,Tobias7617,Richmond5051,AllenSodiumHalide2252,ShultzJPCB585}
In order to systematically investigate the anion effects on the
hydrogen-bonded water molecules at the electrolyte aqueous solution
surfaces, the ssp SFG-VS spectra of the NaF, NaCl, NaBr and NaI slat
solution surfaces were measured by two independent
groups.\cite{Richmond5051,AllenSodiumHalide2252} Their SFG-VS data
on the NaCl, NaBr and NaI aqueous solution surfaces agreed well with
each other. Basically, the intensity of the narrow peak, centered at
3700cm$^{-1}$ and assigned as the stretching vibration mode of `free
O-H', remained almost unchanged for all these salt solution surfaces
upto 2.0M. In the mean time, the intensity of the broad band spectra
of hydrogen bonded water in the broad 3400cm$^{-1}$ band increased
slightly to significantly from that of the neat air/water interface
in the order of NaCl, NaBr, and NaI; while the intensity of the
broad band around 3250cm$^{-1}$ decreased with the same order.
However, the SFG-VS data of the NaF aqueous solution surface from
the two competing groups exhibited significant difference in the
3000-3600cm$^{-1}$ spectral region, while the 3700cm$^{-1}$ remained
almost unchanged from that of the neat air/water interface. Allen
and co-workers observed almost identical ssp SFG-VS spectra for the
neat air/water interface and for the 0.83M NaF aqueous solution
surface, i.e. there was essentially no ion effect of the NaF salt on
the ssp SFG-VS spectra of the interfacial water
molecules.\cite{AllenSodiumHalide2252} However, Richmond and
co-workers observed apparent drop of the SFG spectral intensity in
the 3000-3600cm$^{-1}$ region for the 0.88M NaF aqueous solution
surface from that of the neat air/water
interface.\cite{Richmond5051}

The conclusions based on the different experimental observations in
these two studies were almost entirely
different.\cite{Richmond5051,AllenSodiumHalide2252,Tobias7617} Allen
and co-workers concluded the confirmation of the MD simulation
results on the enhanced adsorption of the more polarizable anions in
the surface water layers.\cite{Tobias7617,TobiasFeatureArticle6361}
They further compared the relative SFG spectral intensity changes
with the respective bulk IR and Raman spectra in the
3000-3600cm$^{-1}$ region, and concluded that the surface water
thickness for the NaBr and NaI aqueous solution was significantly
increased. This was also considered as the evidence to support the
enhanced adsorption of the Br$^{-}$ and I$^{-}$ anions at the
aqueous solution surfaces.

However, Richmond and co-workers suspected that the change of the
SFG intensity in the 3400cm$^{-1}$ for the NaCl, NaBr and NaI
solution surfaces did not necessarily indicate the increase of the
thickness of the surface water layers at the electrolyte solution
surfaces. In the meantime, they also concluded that the drop of the
SFG spectra intensity in the 3000-3600cm$^{-1}$ region for the 0.88M
NaF solution surface was the strong evidence for the presence of the
F$^{-}$ anion in the aqueous solution surface layers, even though
the F$^{-}$ anion concentration in the surface region was less than
that in the bulk. Therefore, the whole paradigm for the surface
enhanced adsorption of the more polarizable anions, i.e. the
Br$^{-}$ and I$^{-}$ anions, was not likely to be supported by their
SFG-VS data. Richmond and co-workers concluded that both the more
polarizable and less polarizable anions, e.g. F$^{-}$ anion, did
present in the vicinity of the sub-surface and influenced the
hydrogen bonding structure in the surface water layers, even without
the enhanced adsorption of any of these anions.\cite{Richmond5051}
Recently, Shen and co-workers employed the phase-sensitive SFG-VS
measurement and studied the ssp SFG-VS spectra of the NaI, HI, HCl
and NaOH aqueous solution surfaces in comparison with the neat
air/water interface.\cite{ShenYRPRL096102,ShenJACS13033} They
confirmed the influence of the interfacial water hydrogen bonding
structure by the I$^{-}$, OH$^{-}$ anions and the hydronium cation
in the previous SFG-VS
studies.\cite{ShultzJPCB585,Allen2007JPCC8814,
RichmondJACS128page14519(2007),ShultzCPL302page157(1999),Richmond5051,AllenSodiumHalide2252,Tobias7617}

To the best of our knowledge, so far there have been no SFG-VS
studies on the cation effects on the hydrogen bonding structure of
the surface water molecules at the electrolyte aqueous solution
surfaces, except for the hydronium cation. There has been no
comparison of the SFG-VS spectra for the NaX (X=f, Cl, Br, or I) and
KX salt aqueous solution surfaces. Our interest on this problem was
the direct result of our recent non-resonant second harmonic
generation (SHG) studies on a series of the alkali halide salt
aqueous solution surfaces.\cite{BianPCCP,BianSubmitted} In these
studies, we demonstrated that the non-resonant SHG signal was
originated only from the electronic responses of the hydrogen bonded
water molecules with the orientationally ordered structures at these
electrolyte aqueous surfaces.\cite{ZhangWKJCP224713} In the SFG-VS
measurement, surface water molecules with different hydrogen bonding
structures exhibit clearly different vibrational spectral features.
However, the non-resonant SHG signal from these surfaces is the
measurement of the whole water surface layer without discrimination
on the surface water molecules in different hydrogen-bonding
configurations. Therefore, non-resonant SHG is a much better tool to
quantify the relative increase of the average thickness of the total
surface water layers than
SFG-VS.\cite{BianPCCP,AllenSodiumHalide2252} With the quantitative
polarization analysis on the non-resonant SHG signals from the
surfaces of the NaF, NaCl, NaBr, KF, KCl, and KBr aqueous solutions
with different bulk concentrations, the relative orientational
parameters as well as the relative changes of the thickness of the
surface water layers for these solutions were obtained
quantitatively.\cite{HongfeiPCCP,BianPCCP,BianSubmitted} In these
studies, we arrived at two main conclusions. The first conclusion
was that the average thickness of the surface water layers increases
with the bulk electrolyte concentration almost linearly in the
following order: KBr $>$ NaBr $>$ KCl $>$ NaCl $\sim$ NaF $>$ KF.
These increases of the average thickness of surface water layers
were less than 35$\%$ of that for the neat air/water interface even
when the bulk salt concentration was 5M for the NaBr and 4M for the
KBr. The second conclusion was that the average orientation
parameter of the water molecules at these surfaces shifts slightly
with the increase of the bulk electrolyte concentration and the
shift was in opposite directions for the sodium and potassium salt
solution surfaces.

These observations explicitly indicated that the Na$^{+}$ and
K$^{+}$ cations can specifically influence the hydrogen bonding
structure of the water molecules at the solution surfaces. From
general chemistry textbooks we know that the saturation solubilities
for NaF and KF salts in water are significantly different at the
room temperatue. In fact, at 20$^{\circ}$C, the saturation
concentration for NaF is only 0.98M (4.13g NaF per 100g water),
while the saturation concentration in KF is 15.4M (89.8g KF per 100g
water).\cite{CRCHandBook} The difference is more than 10 fold. These
facts suggested that the interactions of the Na$^{+}$ and K$^{+}$
cations with water and the F$^{-}$ anion in the aqueous phase can be
significantly different. The vestige of such differences may also be
presented in the hydrogen bonding structure of the surface water
molecules for the aqueous solutions of the NaF and KF salts.
Apparently, SFG-VS is the right tool to reveal these differences in
the aqueous solution surfaces.\cite{ShenYRJPCB3292}

\begin{figure}[h]
\begin{center}
\includegraphics[width=7.5cm]{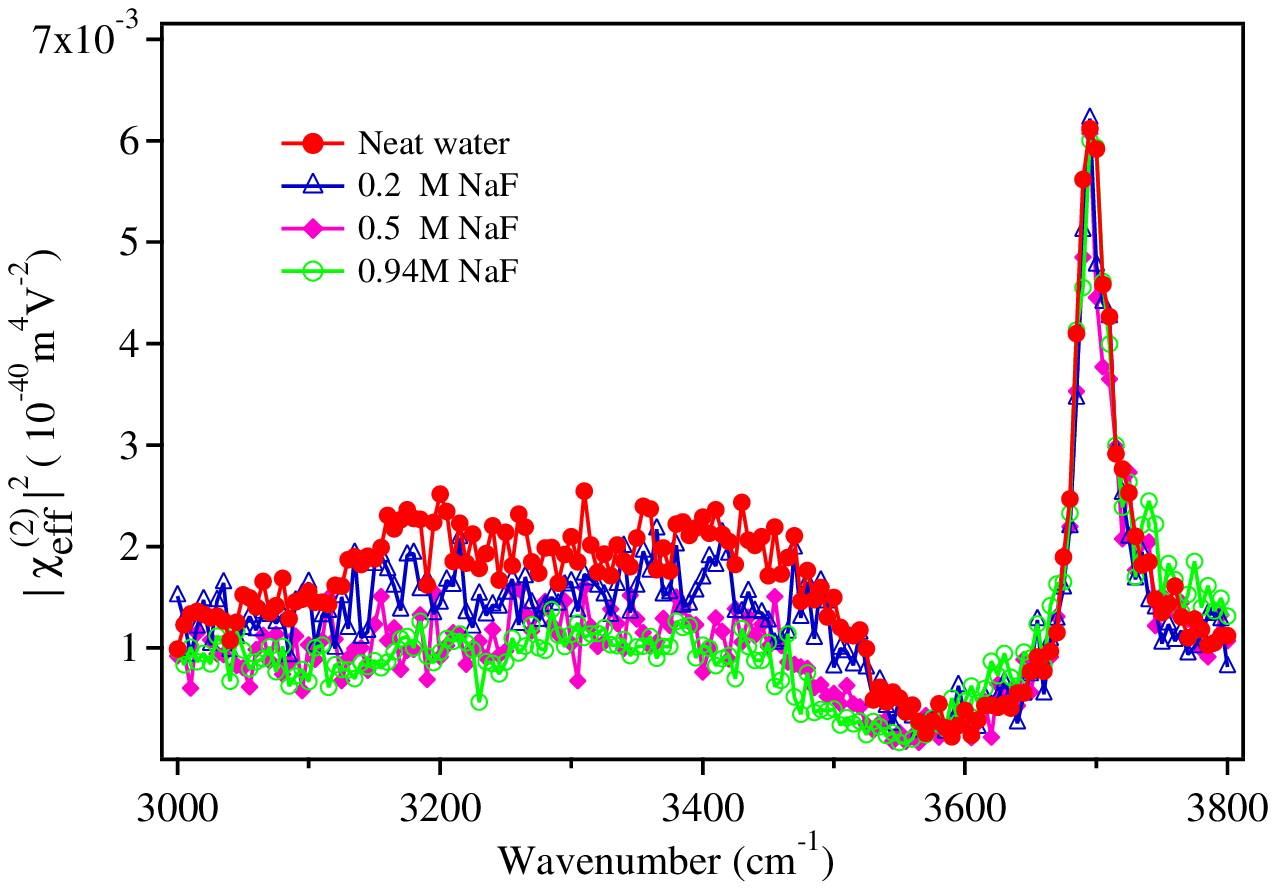}
\\
\vspace{0.3cm}
\includegraphics[width=7.5cm]{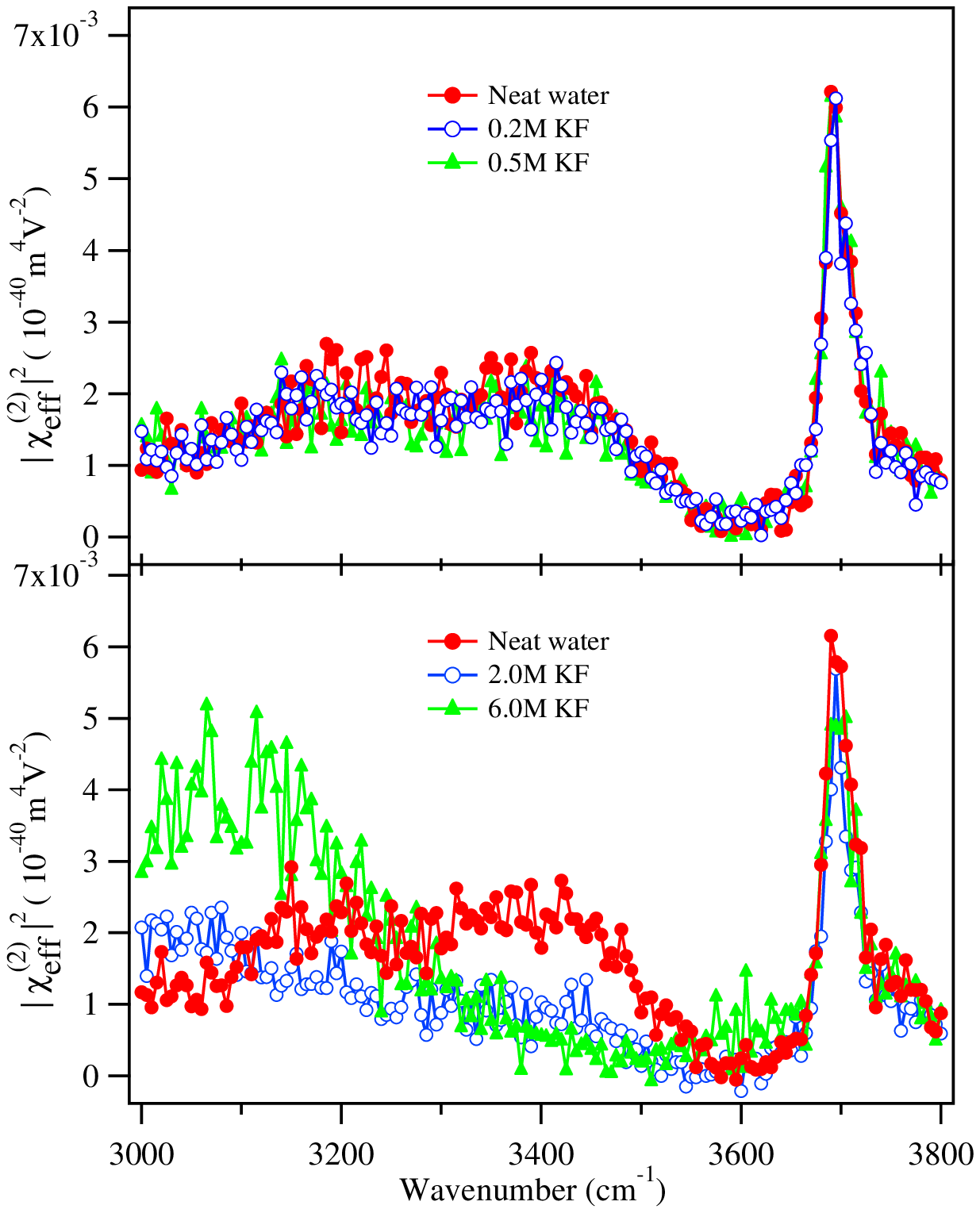}
\caption{SSP SFG-VS spectra of the NaF and KF salt aqueous solution
surfaces at different bulk electrolyte concentrations. Each panel
presents the data measured on the same day, as accompanied by the
SSP SFG-VS spectra of the neat air/water interface measured on that
day. Clearly the salt effects on the SFG spectra in the
3000-3600cm$^{-1}$ hydrogen-bonded water spectral region were quite
different for the NaF and KF
salts.}\label{NaFKFConcentrationDependenceSFGSpectra}
\end{center}
\end{figure}

\begin{figure}[h]
\begin{center}
\includegraphics[width=7.5cm]{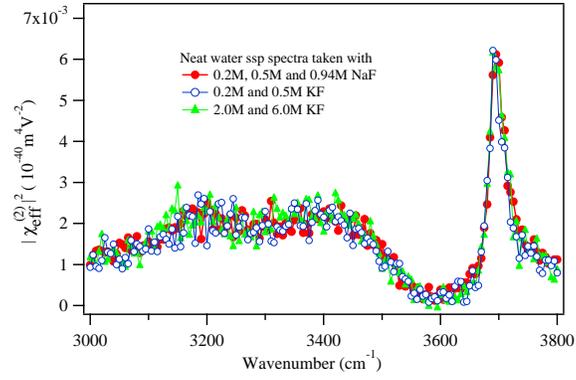}
\caption{Comparison of the three SSP SFG-VS spectra of the neat
air/water interface in the Figure
\ref{NaFKFConcentrationDependenceSFGSpectra}. These spectra were
measured in different days and they agreed with each other
quantitatively. Such good reproducibility of the air/water ssp
SFG-VS spectra suggested that the differences in SFG spectra for the
0.2M and 0.5M NaF and KF solution surfaces as shown in the Figure
\ref{NaFKFConcentrationDependenceSFGSpectra} were real differences
rather than possible experimental
errors.}\label{AirWaterSFGSpectraComparison}
\end{center}
\end{figure}

The ssp SFG-VS spectra in the 3000-3800$cm^{-1}$ region of the neat
air/water interface, and the surfaces of the NaF and KF aqueous
solution with different bulk salt concentrations are presented in
the Figure \ref{NaFKFConcentrationDependenceSFGSpectra} and Figure
\ref{AirWaterSFGSpectraComparison}. The details of the SFG-VS
experiment with a picosecond SFG-VS spectrometer (EKSPLA, Lithuania)
were described previously.\cite{GanWeiJCP114705} The experiments
were conducted in the co-propagation configuration with the visible
beam incident angle at 45$^{\circ}$, the IR beam incident angle at
55$^{\circ}$, and with the SFG signal detection in the reflection
geometry. The SFG signal was normalized with a thick z-cut quartz ,
and the normalization procedures were described
previously.\cite{ShenYRPRL4799,GanWeiJCP114705} The NaF (ACROS and
Sigma-Aldrich, ACS reagent grade with $99\% +$ purity) and KF
(Sigma-Aldrich, ACS reagent grade with $99\% +$ purity) salt samples
were treated with the previously reported purify procedures, and the
possible organic contaminations at the solution surfaces were
monitored with the SHG fluctuation and SFG-VS
spectra.\cite{BianPCCP} The liquid water used in the experiment was
purified with a Millipore Simplicity 185
($18.2{M}{\Omega}{\cdot}cm$) from double distilled water.

The four ssp SFG-VS spectra in the 3000-3800cm$^{-1}$ region as
shown in the top panel of Figure
\ref{NaFKFConcentrationDependenceSFGSpectra} correspond to the neat
air/water interface, the 0.2M, the 0.5M and the 0.94M NaF aqueous
solution surfaces. The apparent concentration dependence of the SFG
signal intensity in the 3000-3600cm$^{-1}$ agreed quantitatively
with the results for the 0.88M NaF solution as reported by Richmond
and co-workers previously.\cite{Richmond5051} These experiments were
conducted with the purified NaF salt sample solutions with special
care.\cite{BianPCCP} The NaF sample from different distributors were
also tested. Since the narrow `free O-H' peak at 3700cm$^{-1}$
showed nearly no change, the NaF solution surface should be free
from any observable organic contaminations. Therefore, the NaF
effects on the spectra in the 3000-3600cm$^{-1}$ region were
evident. Therefore, the observed bulk concentration dependence for
the NaF solution surface was quite firmly established. In the Figure
\ref{AirWaterSFGSpectraComparison} we presented the three spectra of
the neat air/water interface measured in three different days. The
reproducibility of these spectra was remarkable, and the same
quality of overlap can also be achieved with some of the most
reliable spectra of the neat air/water interface in the
literatures.\cite{ShenYRPRL4799,Richmond5051,AllenSodiumHalide2252}

The concentration dependent spectra for the KF aqueous solution
surfaces are presented in the bottom two panels in the Figure
\ref{NaFKFConcentrationDependenceSFGSpectra}. The narrow
3700cm$^{-1}$ peak also remained unchanged for the different bulk KF
concentrations, and the concentration dependent changes of the
3000-3600cm$^{-1}$ for the KF solution surfaces showed no
resemblance to those for the NaF solution surfaces at various
concentrations. Unlike the NaF, for lower concentration KF
solutions, the surface SFG spectra remained almost unchanged from
the SFG spectra of the neat air/water interface in the whole
3000-3800cm$^{-1}$. Nevertheless, at higher KF concentrations of
2.0M and 6.0M, the spectral intensity of the broad 3400cm$^{-1}$
band dropped significantly from that in the SFG spectra of the neat
air/water surface; while the intensity for the broad band below
3200cm$^{-1}$ raised above that in the SFG spectra of the neat
air/water interface. In contrast, both the whole broad band in the
3000-3600cm$^{-1}$ region in the SFG spectra of the NaF solution
surfaces, as shown in the top panel in the Figure
\ref{NaFKFConcentrationDependenceSFGSpectra}, dropped significantly
with the increase of the bulk NaF concentration in comparison to the
spectra intensity of the neat air/water interface in the same
spectral region.

Considering the fact that the solubility of the KF salt in water is
more than 10 times bigger than that of the NaF salt, one may expect
that there are less ions, anions or cations, in the vicinity of the
0.2M and 0.5M KF solution surfaces than in the vicinity of the
surfaces of the NaF solutions with the same bulk concentrations.
However, the striking differences in the SFG spectra features of the
NaF and KF solution surfaces suggested not only the presence of
different quantity of ions at their respective surfaces, but also
specific cation effects of the two electrolytes on the surface water
molecules. Such ion effects certainly can not be explained as the
anion effect since both the NaF and KF solutions consists the common
F$^{-}$ anion. The significantly different Na$^{+}$ and K$^{+}$
cation effects as observed here neither can be easily accommodated
into the classical picture of the electrolyte aqueous solution
surfaces,\cite{OnsagerJCP528,MarkinJPCB11810} nor can be easily
explained with the currently prevalent
paradigm.\cite{TobiasChemRev,DangChemRev,TobiasFeatureArticle6361}

If the classical picture was correct, both the anion and cation
should be repelled from the aqueous surface. Then the observed anion
and cation effects can only be chiefly due to the long-range
interactions between the surface repelled cations and the oriented
water dipoles at the solution surface. According to this picture, it
shall be hard to imagine how such long-range interactions can be so
different for the Na$^{+}$ and the K$^{+}$ cations, since people
already knew that the solvation of the Na$^{+}$ and the K$^{+}$
cations hardly go beyond the second solvation
shell.\cite{LisyJCP8555,LisyJCP024319} On the other hand, in the
currently prevalent paradigm, the adsorption of the more polarizable
anions to the aqueous solution surfaces should be enhanced, but not
the less polarizable anions, such as the F$^{-}$, and the cations,
such as the Na$^{+}$ and the K$^{+}$ cations. Then the difficulties
shall be the same as in the classical picture on how the
significantly different Na$^{+}$ and the K$^{+}$ cation effects are
originated.

Richmond and co-workers observed the change of the SFG spectra for
the NaF aqueous solution surfaces,\cite{Richmond5051} which is
further confirmed in our measurement as reported here. They
concluded that even though the F$^{-}$ anion concentration was
`diminished' in the top surface water layers, the presence of
F$^{-}$ anion in the surface region can still affect the hydrogen
bonding structure of the water molecules in the surface region in
different ways from the larger and more polarizable Br$^{-}$ and
I$^{-}$ anions. According to the same reasoning as Richmond and
co-workers proposed for the F$^{-}$ anion, the presence of the
Na$^{+}$ and the K$^{+}$ cations in the surface layers, even though
diminished from the bulk concentration, can also be expected.
Indeed, in addition to confirming the results on the NaF in their
paper,\cite{Richmond5051} we also observed the cation effects for
the Na$^{+}$ and the K$^{+}$ cations, which can further support the
reasoning proposed by Richmond and co-workers. This latter
observation also indicated that both the Na$^{+}$ cation and the
F$^{-}$ anion were responsible for the observed electrolyte effects
on the water molecules at the NaF solution surface. It can not be
attributed to the F$^{-}$ anion alone, and it is also possible that
ion pairing may play roles.\cite{BianSubmitted} This certainly
provides a new aspect to the understanding of the ions at the
air/water interface. No matter how, since all the SFG-VS
measurements we have conduced are measurements on the surface water
molecules instead of the ions, it is clear to us that neither the
anion effects nor the cation effects on the SFG spectra can be
simply used as the evidences for the enhanced adsorption of either
of the ions at the electrolyte aqueous surface.

In conclusion, the SFG-VS and non-resonant SHG experimental
observations on the water molecules at the solution surfaces of the
various alkali halide salts have demonstrated clear cation effects
at the electrolyte solution surfaces. More experimental data on
different electrolyte systems are required to demonstrate the
various possibilities of the cation effects, and also the ion
pairing effects on the water molecules at the electrolyte solution
surfaces. Based on the observed cation effects of the NaF and KF, we
are now in the process to investigate more NaX/KX salt pairs, as
well as other salts. These observations may provide important clues
to the understandings on the fundamental phenomenon of ions at the
air/water interface.

\vspace{0.3cm}

\noindent\textbf{Acknowledgment:} RRF thanks the technical
assistance from Feng Wang and Zhi Huang. HFW thanks the support by
the Natural Science Foundation of China (NSFC, No.20425309,
No.20533070, No. 20773143) and the Ministry of Science and
Technology of China (MOST No. 2007CB815205). YG thanks the support
by the Natural Science Foundation of China (NSFC, No.20673122).

\clearpage
\begin{list}{}{\leftmargin 2cm \labelwidth 1.5cm \labelsep 0.5cm}

\item[\bf Fig. 1] SSP SFG-VS spectra of the NaF and KF salt aqueous solution
surfaces at different bulk electrolyte concentrations. Each panel
presents the data measured on the same day, as accompanied by the
SSP SFG-VS spectra of the neat air/water interface measured on that
day. Clearly the salt effects on the SFG spectra in the
3000-3600cm$^{-1}$ hydrogen-bonded water spectral region were quite
different for the NaF and KF salts.

\item[\bf Fig. 2] Comparison of the three SSP SFG-VS spectra of the neat
air/water interface in the Figure
\ref{NaFKFConcentrationDependenceSFGSpectra}. These spectra were
measured in different days and they agreed with each other
quantitatively. Such good reproducibility of the air/water ssp
SFG-VS spectra suggested that the differences in SFG spectra for the
0.2M and 0.5M NaF and KF solution surfaces as shown in the Figure
\ref{NaFKFConcentrationDependenceSFGSpectra} were real differences
rather than possible experimental errors.

\end{list}

\clearpage

\end{document}